\documentclass[12pt]{iopart}

\usepackage{epsfig}

\begin{document}

\title{Electronic phase diagram in the new BiS$_{2}$-based Sr$_{1-x}$La$_{x}$FBiS$_{2}$ system}

\author{ Yuke Li$^{1,2}$ \footnote[1]{Electronic address: yklee@hznu.edu.cn}, Xi Lin$^{1}$, Lin Li$^{1}$, Nan Zhou$^{1}$, Xiaofeng Xu$^{1}$ \footnote[2]{Electronic address: xiaofeng.xu@hznu.edu.cn}, Chao Cao$^{1}$, Jianhui Dai$^{1}$, Li Zhang$^{3}$, Yongkang Luo$^{2}$, Wenhe Jiao$^{2}$, Qian Tao$^{2}$, Guanghan Cao$^{2}$ and Zhuan Xu$^{2}$}

\address{$^{1}$Department of Physics and Hangzhou Key Laboratory of Quantum Matters, Hangzhou Normal University, Hangzhou 310036, China}
\address{$^2$State Key Lab of Silicon Materials and Department of Physics, Zhejiang University, Hangzhou 310027, China}
\address{$^3$Department of Physics, China Jiliang University, Hangzhou 310018, China}

\date{\today}

\begin{abstract}
In this paper, we systematically study the electron doping effect in a new BiS$_{2}$-based system
Sr$_{1-x}$La$_{x}$FBiS$_{2}$(0 $\leq x \leq 0.7)$ through multiple techniques of X-ray
diffraction, electrical transport, magnetic susceptibility, and Hall effect measurements. The
parent compound SrFBiS$_{2}$ is found to possess a semiconducting-like ground state, with thermally
activation energy $E_g$ $\sim$ 38 meV. By the partial substitution of La for
Sr, superconductivity emerges when $x >$ 0.3, reaching its maximal superconducting transition
temperature \emph{T$_{c}$} $\sim$ 3.5 K at $x =$ 0.55.
In the normal state of superconducting samples, it is clearly seen that there exists a crossover from metallic to semiconducting state below a temperature $T_{min}$, which shifts to lower temperatures with increasing La content. Based on these measurements, the
associated electronic phase diagram of Sr$_{1-x}$La$_{x}$FBiS$_{2}$ system has thus been
established.

\end{abstract}

\maketitle

\section{\label{sec:level1}Introduction}
In the last couple of decades, the discovery of a series of new superconducting compounds with
relatively low transition temperatures \emph{T$_{c}$} yet displaying some unique properties, has
triggered sustained research interest in the community of condensed matter
physics\cite{SrRuO,NaCO,LaFeP}. In particular, these compounds in some cases have led to the
unveiling of entire new families of surperconducting materials\cite{LaFeP,Hosono,Chen-Sm} with
$T_c$ far above the Bardeen-Cooper-Schrieffer (BCS) limit\cite{BCS}. Interestingly, with few exceptions, these compounds
possess a layered crystal structure, and exhibit exotic superconducting properties and
complex physical features.


Very recently, a novel superconductor Bi$_{4}$O$_{4}$S$_{3}$ with \emph{T$_{c}$} of 8.6 K has been
reported\cite{BOS,BOS2}, which consists of a layered crystal structure built from the stacking of
Bi$_{2}$S$_{4}$ superconducting layers and Bi$_{4}$O$_{4}$(SO$_{4}$)$_{1-x}$ block layers.
Immediately after this work, several new BiS$_{2}$-based superconductors,
\emph{Ln}O$_{1-x}$F$_{x}$BiS$_{2}$(\emph{Ln}=La, Ce, Pr, Nd)\cite{LaFS,NdFS,LaFS2,CeFS,PrFS}, which
are composed of Bi$_{2}$S$_{4}$ layers and \emph{Ln}$_{2}$O$_{2}$ layers, have thus been
discovered, with \emph{T$_{c}$} as high as $~$10 K. Evidently, these compounds share a common
BiS$_{2}$ layer, which serves as the basic building blocks of this new superconducting family. This
feature is reminiscent of the situation encountered in the cuprate and pnictides superconductors,
in which superconductivity arises predominantly from the CuO$_{2}$ planes and Fe$_{2}$Pn$_{2}$
layers, respectively. However these BiS$_{2}$-based layered superconductors display a wealth of
similarities to the iron pnictides, the differences are also manifest. For instance, the parent
compound of LnFeAsO shows an antiferromagnetic (AFM) transition and/or a structural phase
transition\cite{DaiPC} and superconductivity emerges from the suppression of this AFM order by
chemical doping or pressure. In contrast, the parent compound \emph{Ln}OBiS$_{2}$ displays no
magnetic/structural transition, implying that magnetism is of less relevance to superconductivity
in BiS$_{2}$-based systems. On the other hand, it is noted that superconductivity is in close
proximity to an insulating normal state for the BiS$_{2}$-based compounds \cite{CeFS}. Yet for the discovered
superconducting materials, superconductivity grows from a metallic normal state.


More recently, a new BiS$_{2}$ based superconductor Sr$_{1-x}$La$_{x}$FBiS$_{2}$ ($x$=0.5) has been
synthesized and studied\cite{LiSrF,SrF}. This compound is iso-structural to LaOBiS$_{2}$, where the
[Ln$_{2}$O$_{2}$]$^{2-}$ layer is replaced by iso-charged [Sr$_{2}$F$_{2}$]$^{2-}$ block, in
analogy to the case of LaOFeAs and SrFFeAs. The parent compound SrFBiS$_{2}$ is a semiconductor, and the 50$\%$ substitution of La for Sr induces superconductivity with \emph{T$_{c}$} of 2.8 K.

To date, most studies about BiS$_{2}$-based systems have focused on their electronic
structures\cite{es}, superconducting transition temperature\cite{SCT} and the pairing
symmetry\cite{HJP,Yildirim}, with only few reports on the study of their electronic phase diagrams.
In this paper, we report the successful synthesis and the detailed characterization of a series of
La-doped Sr$_{1-x}$La$_{x}$FBiS$_{2}$ samples. We find that superconductivity is successfully
induced by La doping when $x >$ 0.3, with maximal $T_c$ of 3.5 K at $x =$ 0.55, above which $T_c$
shows less doping dependence. Interestingly, for the superconducting samples, their normal state
undergoes a metal to semiconductor transition/crossover at \emph{T$_{min}$}, which shifts to lower
temperatures with increasing La content. For all superconducting samples studied, a strong
diamagnetic signal was observed, confirming the bulk superconductivity. As a consequence, we
established the electronic phase diagram of this Sr$_{1-x}$La$_{x}$FBiS$_{2}$ system based on our
experimental data.

\section{\label{sec:level1}Experiment}

The polycrystalline samples of Sr$_{1-x}$La$_{x}$FBiS$_{2}$ (0$\leq$x$\leq$ 0.7) used in this study
were synthesized by two-step solid state reaction method. The detailed synthesis procedures can be
found in our previous report\cite{LiSrF}. Crystal structure characterization was performed by
powder X-ray diffraction (XRD) at room temperature using a D/Max-rA diffractometer with
CuK$_{\alpha}$ radiation and a graphite monochromator. Lattice parameters were obtained by Rietveld
refinements. The electrical resistivity was measured with a standard four-terminal method between
300 K to 0.4 K in a commercial Quantum Design PPMS-9 system with a $^{3}$He refrigeration insert.
The Hall effect measurements were also performed in this system. The temperature dependence of d.c.
magnetization was measured by a Quantum Design MPMS-5.

\section{\label{sec:level1}Results and Discussion}

Figure 1 (a) shows the room temperature powder XRD patterns of the Sr$_{1-x}$La$_{x}$FBiS$_{2}$  samples. The main
diffraction peaks of these samples can be well indexed based on a tetragonal cell structure with
the P4/nmm space group, expect for some extra minor peaks arising from the possible impurity phase
of Bi$_{2}$S$_{3}$ (note that $x =$0.4 sample shows a single phase only). Fig. 1(b) displays the
(102) and (004) diffraction peaks on an enlarged scale. It is clearly seen that the (004) peak for
La-doped samples shifts systematically towards higher 2$\theta$ angles with respect to that in the
parent compound ($x=$ 0), while the (102) peak exhibits much less doping dependence. This
observation is consistent with the variation of the lattice parameters at room temperature, as shown in Fig. 1(c),
where the $c$-axis decreases quickly as $x$ grows, while the $a$-axis is nearly independent of the
La content. These suggest that La atoms are indeed incorporated into the lattice, resulting in
the decrease in the cell volume. Similar feature was also reported in the case of
LaO$_{1-x}$F$_{x}$BiS$_{2}$\cite{LaFS2}. To obtain the actual content of La impurities, the EDX measurements (not shown here) are performed in a single crystal grain of polycrystalline sample for all La-doped samples. Those results suggest that the real La content in all doped samples is very close to the nominal La concentration.

The temperature evolution of the resistivity $\rho$(T) for all samples studied is summarized in
Fig. 2. The parent compound SrFBiS$_{2}$, which has been broadly studied in the
literature\cite{LiSrF,SrF}, is a semiconductor with room temperature resistivity of $\sim$
5$\times10^{2} $m$\Omega\cdot$cm. This room temperature value is about two orders of magnitude
larger than that of iron-based superconductors with semi-metallicity\cite{Hosono}. No anomaly in
resistivity is observed down to 2 K, contrasting with the prominent kink structure associated with
the AFM phase transition in \emph{Ln}FeAsO systems\cite{DaiPC}. Remarkably, the thermal energy
$E_g$ extracted from the fitting to the thermal activation formula $\rho(T)=\rho_0 \exp(E_g/k_B T)$
at the temperature range from 100 K to 300 K, is about 38.2 meV for the parent compound. When 30\%
of La is introduced, however, the resistivity decreases sharply yet remains semiconducting-like with
its negative $T$-coefficient of the resistivity. Note that the thermal activation energy $E_g$
increases to 58 meV, which was ascribed to the impurity phase of Bi$_{2}$S$_{3}$ with sulfur
deficiency\cite{BiS}. As $x$ further increases to 0.4, resistivity shows semiconducting behavior
before a sharp superconducting transition with $T_{c}^{onset}$ (the onset temperature at which the
resistivity starts to drop) $\sim$ 2.0 K sets up, as clearly seen in the low-$T$ enlarged plot in
Fig. 2(b). This result is comparable to the case of LaO$_{0.5}$F$_{0.5}$BiS$_{2}$, where the normal
state shows semiconducting behavior, and it undergoes a superconducting transition below $~$10
K\cite{LaFS}. The $E_g$ fitted from high temperature decreases from 30 meV for $x = $0.4 to 8.6 meV
for $x = $0.5, suggesting the decrease of gap size due to electron doping. The highest
\emph{$T_{c}^{onset}$} is observed to be 3.5 K at $x =$ 0.55. Interestingly, for the
superconducting samples, there occurs a crossover from a metallic to semiconducting
state as the temperature is lowered below $T_{min}$. While the \emph{T$_{min}$} shifts
progressively towards lower temperatures with increasing La concentration up to $x$=0.7,
superconducting \emph{$T_{c}$} remains almost unchanged above the doping level of 0.55, as observed
in the Fig. 2(b).

To further confirm the bulk nature of the observed superconductivity, we performed the d.c.
magnetic susceptibility measurement with both zero field cooling (ZFC) and field cooling (FC) modes
under 5 Oe magnetic field for the superconducting samples, as depicted in Fig. 3. For all
superconducting samples, strong diamagnetic signals are observed and the \emph{T$_{c}$} values
determined from the magnetic susceptibility are overall consistent well with the resistivity data.
The estimated volume fraction of superconducting shielding from ZFC data is close to 30\%. Note
that the Meissner volume fraction estimated by FC data also exceeds 15\%, which is larger than
the previous report in LaO$_{1-x}$F$_{x}$BiS$_{2}$ system\cite{EPL}.

Fig. 4 shows the temperature dependence of resistivity for $x =$0.55 sample under various magnetic
fields below 5 K. It can be seen that $T_{c}^{onset}$ is about 3.5 K at zero field. With
increasing magnetic fields, $T_{c}^{onset}$ is gradually suppressed and the superconducting
transition becomes obviously broader. Above 0.3 T, $T_{c}^{onset}$ becomes robust in fields and its
value shows nearly no changes at around 2.6 K. For $B =$ 2 T, however, the low temperature resistivity increases rapidly, but a slight drop at 2.6 K is still detectable. The inset shows the temperature dependence of upper critical field $\mu_0$$H_c(T)$ below 5 K, determined by using 99\% normal state resistivity criterion. The $\mu_0$$H_c(T)$ is nearly linear with temperature decreasing at low field but becomes constant as $H$ is over 0.4 T. This may be ascribed to the residual Cooper pairs existing in the BiS$_2$-based
system, where similar behaviors were also reported in the Bi$_{4}$O$_{4}$S$_{3}$
system\cite{WenHH}. According to the conventional one band WHH formula\cite{WHHF} $\mu_{0}H_{c2}(0) = -0.69T_{c0}(\partial{H_{c2}}/\partial{T})|_{T_{c0}}$,
the upper critical field $\mu_0$$H_{c2}$ estimated is about 1.44 T. Apparently, this result from one band WHH model does not match to our experimental data, indicating the mutilband superconductivity in the Sr$_{1-x}$La$_{x}$FBiS$_{2}$ system.
Recently, the calculations\cite{WangQH} have claimed that in BiS$_{2}$ systems, the triplet pairing interaction is so strong that it may become dominant. Therefore, it would be intriguing to study its low-lying quasiparticle excitations in the present
system\cite{Xu13,Niu13}.


Recently, the first principles calculations\cite{Yildirim} suggested a charge density wave
instability or an enhanced correlation effect in this system. In order to study its normal state
properties and obtain some useful insights into such a putative instability, the Hall effect
measurement was performed. Fig. 5 shows the temperature dependence of $R_{H}$ for the superconducting samples as $x$ $\geq$ 0.4.
The inset gives the magnetic field dependence of the transverse resistivity $\rho_{xy}$ at
various temperatures for the representative $x$ =0.55 sample. In the main panel of Fig. 5, the positive $R_{H}$ throughout the entire temperature range is observed and displays a weak temperature dependence above 50 K for $x$ =0.4 sample, suggesting the dominant hole-carriers regardless of electron doping in this compound. In contrast, as $x$ $\geq$ 0.5, the  $R_{H}$ in the whole temperature
region is negative and indicates that the electron-type charge carriers is dominant for those compounds. Moreover, their $R_{H}$ are seen to be
$T$-independent at high temperatures and drop drastically below 100 K. The sign change of $R_{H}$ from positive to negative may be ascribed to the suddenly topology change of the Fermi Surface according to the calculations\cite{Usui,Mart}. On the other hand, as shown
in the inset of Fig. 5, the magnetic field dependence of $\rho_{xy}$ shows weak nonlinear behaviors
at low temperatures for $x =$0.55. The results are consistent with the multi-band effects in the present
system, similar to the case of CeOFBiS$_{2}$\cite{CeFS}. However, another possibility of the spin
(charge) density wave (SDW/CDW) formation\cite{Oak} can not be ruled out at this stage, as the
similar sharp drop in $R_{H}$ was also observed in the iron-based superconductors below AFM
transition\cite{Oak2}.

On the basis of the above experimental findings, the electronic phase diagram of
Sr$_{1-x}$La$_{x}$FBiS$_{2}$ system is mapped out in Fig. 6. The parent compound SrFBiS$_{2}$ is a
semiconductor. With La doping on Sr sites, superconductivity emerges as $x$ $\geq$ 0.3 and it
reaches a maximal $T_c$ $\sim$ 3.5 K at $x$=0.55. Meanwhile, a crossover from metallic to
semiconducting-like behavior is observed in the normal state of the superconducting samples, with
the resistivity minimum $T_{min}$ shifting progressively towards lower temperatures with doping.

The above phase diagram of Sr$_{1-x}$La$_{x}$FBiS$_{2}$ system distinguishes itself from those of
\emph{Ln}O$_{1-x}$F$_{x}$BiS$_{2}$ systems in several aspects. First, the parent compound of the
former is a semiconductor\cite{LiSrF} while the latter is a bad metal\cite{CeFS}. Second, by
electron doping, the resistivity decreases with the doping concentration in
Sr$_{1-x}$La$_{x}$FBiS$_{2}$, which is distinct from \emph{Ln}O$_{1-x}$F$_{x}$BiS$_{2}$\cite{CeFS}
and La$_{1-x}$M$_{x}$OBiS$_{2}$ (M=Ti, Zr, Hf, Th) systems\cite{LaM}, where resistivity shows the
opposite trend with the electron concentration. Third, no metal to semiconductor
transition/crossover has been observed in \emph{Ln}O$_{1-x}$F$_{x}$BiS$_{2}$ and
La$_{1-x}$M$_{x}$OBiS$_{2}$ systems.

\section{\label{sec:level1}Conclusion}

In summary, we have successfully synthesized a series of BiS$_{2}$-based
Sr$_{1-x}$La$_{x}$FBiS$_{2}$ polycrystalline samples. Through the measurements of resistivity and
magnetic susceptibility, the parent compound was found to be semiconducting-like with an thermal activation energy $E_g$ $\sim$ 38 meV. Via the partial La doping on Sr sites, however, the resistivity decreases
sharply, and ultimately superconductivity emerges as $x >$ 0.3 and it reaches \emph{T$_{c}$} of
3.5 K at the optimal level of $x =$ 0.55. The superconducting samples undergo a metal to
semiconductor transition/crossover below \emph{T$_{min}$}, which is seen to be gradually suppressed
with further doping whilst $T_c$ remains nearly unchanged above the optimal doping. Hall effect
measurements for the La-doped samples confirm that the sign change of $R_{H}$ from positive to negative may be ascribed to the suddenly topology change of the Fermi Surface.  A drop in \emph{R$_{H}$} and the non-linear transverse resistivity $\rho_{xy}$ with field are associated with the multi-band effect or a CDW instability. According to these measurements, an electronic phase diagram is thus
established.

\section*{Acknowledgments}
Y. K. Li would like to thank Bin Chen, Jinhu Yang, Hangdong Wang for useful discussions, and
Xuxin Yang, Quanlin Ye for collaborative support. This work is supported by the National Basic
Research Program of China (Grant No. 2011CBA00103 and 2012CB821404), NSFC (Grant No. 11174247,
11104053, 61376094, 11104051).

\pagebreak[4]

\begin{figure}
\includegraphics[width=16cm]{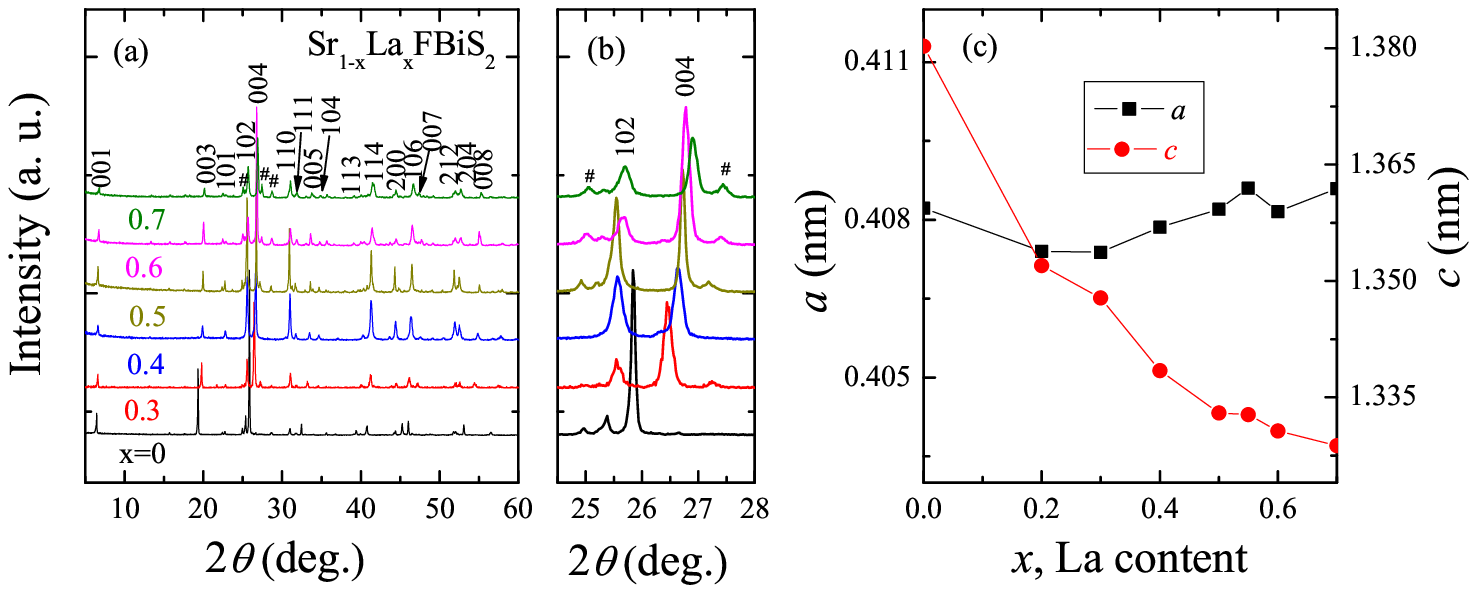}
\caption{\label{Fig.1} (color online). Crystal structure of Sr$_{1-x}$La$_{x}$FBiS$_{2}$ (a)Room temperature powder
X-ray diffraction patterns of Sr$_{1-x}$La$_{x}$FBiS$_{2}$ (0$\leq$x$\leq$0.7) samples. The \# peak
positions designate the impurity phase of Bi$_{2}$S$_{3}$. (b) An enlarged plot of the XRD
diffraction peaks around (102) and (004). (c) Lattice parameter as a function of La content $x$.}
\end{figure}

\begin{figure}
\includegraphics[width=8cm]{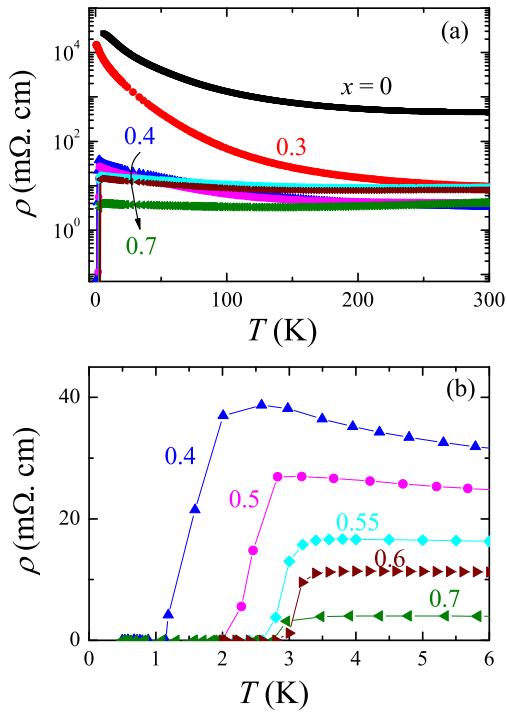}
\caption{\label{Fig.2} (color online). (a) Temperature dependence of resistivity $\rho$($T$) for
the Sr$_{1-x}$La$_{x}$FBiS$_{2}$ (0$\leq$x$\leq$ 0.7) samples. (b) A close-up view of the
resistivity around $T_{c}$ for the superconducting samples.}
\end{figure}

\begin{figure}
\includegraphics[width=8cm]{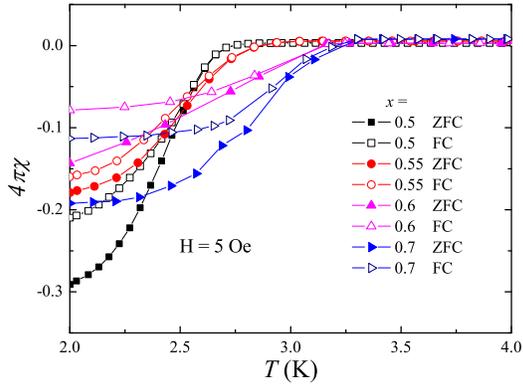}
\caption{\label{Fig.2} (color online). Temperature dependence of the magnetic susceptibility under
5 Oe magnetic field with ZFC(Solid) and FC(Open) modes for $x =$ 0.5, 0.55, 0.6, 0.7 samples. The Meissner
volume fraction for those samples is over 15\%, confirming the bulk superconductivity.}
\end{figure}

\begin{figure}
\includegraphics[width=8cm]{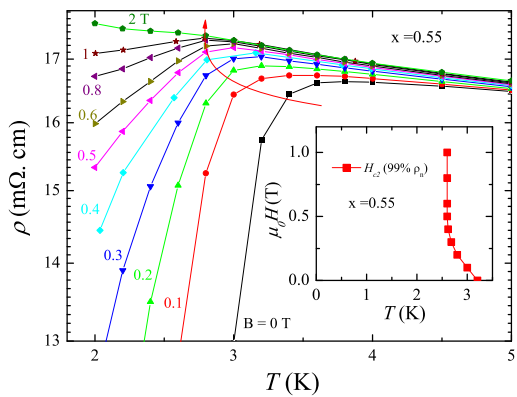}
\caption{\label{Fig.4}(color online) Temperature dependence of resistivity below 5 K under several
magnetic fields for $x =$ 0.55 sample. The inset: $\mu_0\emph{H(T)}$ as a function of temperature.}
\end{figure}

\begin{figure}
\includegraphics[width=8cm]{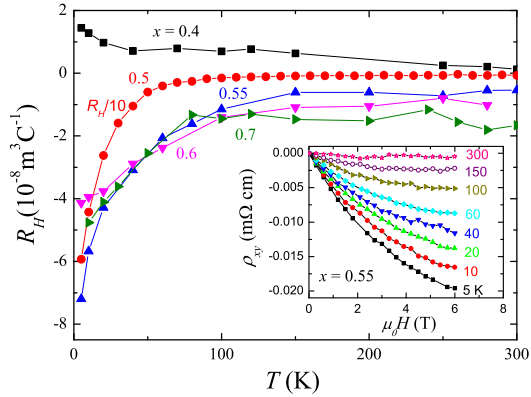}
\caption{\label{Fig.3}(color online) Temperature dependence of Hall coefficient $R_{H}$ measured at
5 T for the different La doping samples. $R_{H}$ shows weak temperature dependence at the high temperature and drops at low temperature. The inset gives the Hall resistivity $\rho_{xy}$ as a function of fields at several representative temperature as $x =$ 0.55. The $\rho_{xy}$ vs. magnetic field with weak nonlinear behaviors suggests the multi-band effect in this system.}
\end{figure}

\begin{figure}
\includegraphics[width=8cm]{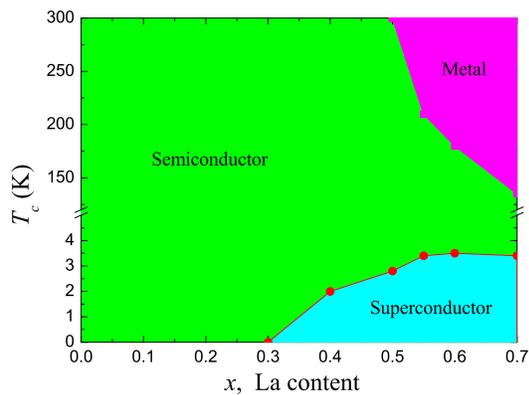}
\caption{\label{Fig.4}(color online) The electronic phase diagram extracted from the resistivity
measurements.}
\end{figure}

\end{document}